\documentclass[aps,prstab,reprint,groupedaddress,amsmath,amssymb,showpacs,showkeys,floatfix,longbibliography]{revtex4-1}

\usepackage{graphicx}

\usepackage{dcolumn}
\usepackage{bm}

\begin{document}

\title{Quantum and classical dynamics of neutron in a magnetic field
\thanks{This work has been supported by Russian Science Foundation ...}
}

\author{ A.~Bogomyagkov }
  \email{A.V.Bogomyagkov@inp.nsk.su}
  \affiliation{Budker Institute of Nuclear Physics SB RAS, Novosibirsk 630090, Russia}
\author{ V.~Druzhinin }
  \affiliation{Budker Institute of Nuclear Physics SB RAS, Novosibirsk 630090, Russia}
\author{ E.~Levichev }
  \affiliation{Budker Institute of Nuclear Physics SB RAS, Novosibirsk 630090, Russia}
\author{ A.~Milstein }
  \affiliation{Budker Institute of Nuclear Physics SB RAS, Novosibirsk 630090, Russia}
\author{ I.~Okunev}
  \affiliation{Budker Institute of Nuclear Physics SB RAS, Novosibirsk 630090, Russia}
\author{ S.~Taskaev }
  \affiliation{Budker Institute of Nuclear Physics SB RAS, Novosibirsk 630090, Russia}

\date{\today}

\begin{abstract}
The paper presents solution of quantum problem of neutron propagation in the magnetic field with multipole field expansion. Rigorous solution of the Pauli equation for neutron reveals existence of two solutions, finite and infinite, for any multipole configuration. As an example, we present detailed study of neutron motion in quadrupole and sextupole magnets. Our predictions agree with the results of Stern-Gerlach experiment for neutrons. To verify existence of finite and infinite motion, we discuss an experiment which could be performed in the Budker Insitute of Nuclear Physics using existing equipment. We conclude with considerations of neutron storage ring with straight section and discrete magnets focusing the beam.
\end{abstract}

\maketitle

\section{Introduction}
Due to absence of an electrical charge, the only way to control neutron motion is using the interaction of magnetic moment and the magnetic field with high gradient. Feasibility of this method was demonstrated in famous Stern-Gerlach experiment, providing the first direct experimental evidence of spin existence \cite{Stern-Gerlach:2016,Zelevinsky:2011zza}. The effect of beam splitting, which we call Stern-Gerlach effect (SGE), with neutrons was reported in \cite{Sherwood:1954,PTA1965en} in the experimental setup similar to the original Stern-Gerlach experiment.

In many textbooks \cite{Platt:1992,Hannout:1998,Rodriguez:2017,Romero:2024} SGE is explained using magnetic field $B_x(x)$ and interaction operator 
\begin{equation}
H=-\mu \boldsymbol{\sigma} \cdot\mathbf{B}=-\mu\sigma_xB_x(x), 
\end{equation}
where $\boldsymbol{\sigma}$ are Pauli matrices, $\mu$ is a magnetic moment of spin-$1/2$ particle. Since the spin projection on axis $x$ is $\pm1/2$, then the corresponding force is
\begin{equation}
F_x=\pm\mu \partial B_x(x)/\partial x. 
\end{equation}
As a result, the beam of unpolarized particles splits into two with distinct spin projections. In particular, magnetic field is chosen as $B_x(x)=B_0+Gx$ (shifted skew quadrupole), where $B_0$ and $G$ are constants. However, such field violates Maxwell equation $\operatorname{div} \mathbf{B}=0$, and it is necessary to add a component $B_y(y)=-Gy$, demanding more accurate quantum analysis of SGE, than naive explanation given in the textbooks. 
However, in spite of numerous attempts to solve the problem of SGE correctly \cite{KUGLER1985,Platt:1992,Hannout:1998,Rodriguez:2017,Romero:2024}, none has found a correct solution. 

Note that the problem possesses an internal symmetry, which becomes obvious with replacement of the variable $x$ with $x'=x+B_0/G$, giving $B(x')=Gx'$ and providing a symmetry between coordinates $x'$ and $y$. Similar symmetry exists for any $2(n+1)$ pole magnet. 

In the present paper, we have solved a quantum problem of neutron motion in $2(n+1)$ pole magnet, yielding a correct transition from quantum description of neutron motion in the magnetic field to classical one.

Using our approach for the case $n=1$, we have explained the experimental results of Stern-Gerlach experiment for neutrons \cite{PTA1965en}. 
We have also revealed correct operating principles of the magnetic trap and the storage ring for ultracold neutrons \cite{KUGLER1985,Paul:1989hn,Anton:1989fh}. A proposal of the strong focusing synchrotron for neutrons was suggested in \cite{Lee:2024wgh}. However, it is based on incorrect explanation of SGE because, as we explained above, the authors used only one magnetic field component thus violating Maxwell equation.

\section{Pauli equation for neutron in $2(n+1)$ pole magnet}
Let's start with solving the stationary Pauli equation (extension of the Schr\"odinger equation for spin-$1/2$ particles in the external electromagnetic field) for neutron in $2(n+1)$ pole magnet. The numbers $n=1$ and $n=2$ describe quadrupole and sextupole, respectively. Magnetic field satisfies Maxwell equations $\operatorname{div} \mathbf{B}=0$ and $\operatorname{rot}\mathbf{B}=0$, while, neglecting the fringe fields, the scalar potential $\Phi$ ($\mathbf{B}=\bm\nabla \Phi$) satisfies 2-d Laplace equation. Cartesian components of the field in harmonic expansion are
\begin{equation}
\label{eq:field-0}
\begin{gathered}
B_x=G_n\rho^n\sin(n\varphi), \mspace{9mu} B_y=G_n\rho^n\cos(n\varphi), \mspace{9mu} B_z=0,
\end{gathered}
\end{equation}
where $\rho=\sqrt{x^2+y^2}$, $\{\rho,\varphi,z\}$ are cylindrical coordinates, and $G_n$ is a constant.
The corresponding stationary Pauli equation reads
\begin{equation}
\label{eq:pauli-1}
\begin{aligned}
E\Psi(\mathbf{r})&=H\Psi(\mathbf{r}), \\
H&=\frac{\mathbf{p}^2}{2M}-\mu \boldsymbol{\sigma} \cdot\mathbf{B}  \\
&  =\frac{\mathbf{p}^2}{2M}-\mu G_n\rho^n[\sin(n\varphi)\sigma_x+\cos(n\varphi)\sigma_x],
\end{aligned}
\end{equation}
where $M$ is the neutron mass with $M c^2=939.565 \text{ MeV}$, $c$ is a speed of light, $\mu=-1.913\mu_N=-6.03\times 10^{-8}\text{eV}/\text{T}$ is the neutron magnetic moment, $\mu_N=3.152\times 10^{-8}\text{eV}/\text{T}$ is the nuclear magneton, $\mathbf{p}=-i\hbar\pmb{\nabla}$ is the momentum operator.

At first, let's show equivalence of the problem of neutron motion in quadrupole lens and in skew quadrupole (used in Stern-Gerlach experiment). We transform the wave function according to $\Psi(\mathbf{r})=\exp(-i\pi s_z/2)\Psi_1(\mathbf{r})$, which rotates spin at angle $\pi/2$ around $z$ axis (the spin operator is $\mathbf{s}=\boldsymbol{\sigma}/2$). Observing that
\begin{equation}
\begin{aligned}
\exp(i\pi \sigma_z/4)\sigma_x\exp(-i\pi \sigma_z/4)&=-\sigma_y, \\
\exp(i\pi \sigma_z/4)\sigma_y\exp(-i\pi \sigma_z/4)&= \sigma_x,
\end{aligned}
\end{equation}
we find that $\Psi_1(\mathbf{r})$ satisfies equation \eqref{eq:pauli-1} with magnetic field $B_x=G_1x$, $B_y=-G_1y$, $B_z=0$. Thus, magnetic field in Stern-Gerlach experiment corresponds to the field of the shifted and rotated quadrupole.

Since magnetic field $\mathbf{B}$ in \eqref{eq:pauli-1} is independent of $z$, we write solution $\Psi(\mathbf{r})$ as
\begin{equation}
\label{eq:solution-1}
\begin{gathered}
\Psi(\mathbf{r})=e^{ikz}\psi(x,y), \\
\psi(x,y)=f_{+}(x,y)\Phi_{+} + f_{-}(x,y)\Phi_{-}, \\
\Phi_{+}=\begin{pmatrix} 1\\ 0 \end{pmatrix},  \mspace{18mu} \Phi_{-}=\begin{pmatrix} 0\\ 1 \end{pmatrix},
\end{gathered}
\end{equation}
where the spin quantization axis is the axis $z$. For functions $f_{\pm}$ we obtain a system of equations
\begin{equation}
\begin{aligned}
\varepsilon f_{+}(x,y)&= \frac{\bm p_\perp^2}{2M}\,f_{+}(x,y)+i\mu\,G_n\,\rho^n\, e^{in\varphi}\,f_-(x,y), \\
\varepsilon f_{-}(x,y)&= \frac{\bm p_\perp^2}{2M}\,f_{-}(x,y)-i\mu\,G_n\,\rho^n\, e^{-in\varphi}\,f_+(x,y),
\end{aligned}
\end{equation}
where $\varepsilon=E-\hbar^2k^2/2M$, $\bm p_\perp^2=p_x^2+p_y^2$.
Using expression
\begin{equation*}
\bm p_\perp^2=-\hbar^2\left[\frac{1}{\rho}\frac{\partial}{\partial \rho}\rho\frac{\partial}{\partial \rho} + \frac{1}{\rho^2}\frac{\partial^2}{\partial \phi^2} \right]
\end{equation*}
and depicting $f_{\pm}$ as 
\begin{equation}
\begin{aligned}
f_+(x,y)= e^{i(m+n/2)\varphi}\,g_+(\rho), \\
f_-(x,y)= e^{i(m-n/2)\varphi}\,g_-(\rho),
\end{aligned}
\end{equation}
where $m$ is integer,
we obtain
\begin{equation}
\begin{aligned}
\varepsilon g_+(\rho)&= \frac{\hbar^2}{2M}\left[-\frac{1}{\rho} \frac{\partial}{\partial \rho}\rho\frac{\partial}{\partial \rho}+ \frac{(m+n/2)^2}{\rho^2}\right] g_+(\rho) \\
&\mspace{216mu}+i\mu G_n \rho^n  g_-(\rho), \\
\varepsilon g_-(\rho)&= \frac{\hbar^2}{2M}\left[-\frac{1}{\rho} \frac{\partial}{\partial \rho}\rho\frac{\partial}{\partial \rho}+ \frac{(m-n/2)^2}{\rho^2}\right] g_-(\rho) \\
&\mspace{216mu} -i\mu G_n \rho^n  g_+(\rho).
\end{aligned}
\end{equation}
Additional transformation 
\begin{align}
g_+(\rho)&=\frac{ \chi_+(\rho)}{\sqrt{\rho}}, &
g_-(\rho)&=\frac{ \chi_-(\rho)}{\sqrt{\rho}}, &
\chi_\pm(0)&=0
\end{align}
will further simplify equations.
For convenience, we introduce dimensionless parameters 
\begin{equation}
\label{eq:dmnless-parameters}
\begin{gathered}
\mathcal{E}=\frac{\varepsilon}{\varepsilon_0} \mspace{18mu} \text{and} \mspace{18mu} \varrho =\frac{\rho}{a_0}, \\
a_0=\left(\frac{\hbar^2}{M |\mu| G_n}\right)^{1/(n+2)}, \quad \varepsilon_0=|\mu| G_n a_0^n.
\end{gathered}
\end{equation}
Noticing, that $\mu=-|\mu|$ we obtain
\begin{equation}
\label{eq:chi}
\begin{aligned}
\mathcal{E} \chi_+(\varrho)&= \frac{1}{2}\left[-\frac{\partial^2}{\partial \varrho^2}+ \frac{(m+n/2)^2-1/4}{\varrho^2}\right] \chi_+(\varrho) \\
&\mspace{234mu}-i\varrho^n  \chi_-(\varrho), \\
\mathcal{E} \chi_-(\varrho)&= \frac{1}{2}\left[-\frac{\partial^2}{\partial \varrho^2}+ \frac{(m-n/2)^2-1/4}{\varrho^2}\right] \chi_-(\rho) \\
&\mspace{234mu}+i\varrho^n \chi_+(\varrho).
\end{aligned}
\end{equation}
Feasible quadrupole gradient $G_1=100$~T/m gives values
\begin{align}
a_0&=2\cdot 10^{-4}\text{ cm}, & \varepsilon_0&=\mu G_1 a_0= 10^{-11} \text{ eV}.
\end{align}
Since $a_0$ and  $\varepsilon_0$ are too small, we need to care about solutions with  $\varepsilon\gg \varepsilon_0$ ($\mathcal{E}\gg 1$), $\rho\gg a_0$ ($\varrho\gg 1$) and $|m|\gg 1$. With $|m|\gg 1$ we can change $((m\pm n/2)^2-1/4)$ to $m^2$ in \eqref{eq:chi}, i.e. changing equations \eqref{eq:chi} to
\begin{equation}
\label{eq:chi-1}
\begin{aligned}
\mathcal{E} \chi_+(\varrho)&= \frac{1}{2}\left[-\frac{\partial^2}{\partial \varrho^2}+ \frac{m^2}{\varrho^2}\right]\chi_+(\varrho)-i\varrho^n \chi_-(\varrho), \\
\mathcal{E} \chi_-(\varrho)&= \frac{1}{2}\left[-\frac{\partial^2}{\partial \varrho^2}+ \frac{m^2}{\varrho^2}\right]\chi_-(\rho)+i\varrho^n \chi_+(\varrho).
\end{aligned}
\end{equation}
This system possesses two solutions. For the first solution 
$\chi_-(\varrho)=i\chi_+(\varrho)=iF_m(\varrho)$  and
\begin{equation}
\label{eq:res-1}
\mathcal{E} F_m(\varrho)= -\frac{1}{2} \frac{\partial^2}{\partial \varrho^2} F_m(\varrho)+\left( \frac{m^2}{2\varrho^2}+\varrho^n\right) F_m(\varrho),
\end{equation}
which describes finite motion in the potential 
\begin{equation}
U_+(\varrho)=\frac{m^2}{2\varrho^2}+\varrho^n.
\end{equation}
For the second solution 
$\chi_-(\varrho)=-i\chi_+(\varrho)=-i\widetilde F_m(\varrho)$  and
\begin{equation}
\label{eq:res-2}
\mathcal{E}\widetilde F_m(\varrho)= -\frac{1}{2}\frac{\partial^2}{\partial \varrho^2} \widetilde F_m(\varrho)+\left(\frac{m^2}{2\varrho^2}-\varrho^n\right) \widetilde F_m(\varrho),
\end{equation}
which describes infinite motion in the potential
\begin{equation}
U_-(\varrho)=\frac{m^2}{2\varrho^2}-\varrho^n.
\end{equation}

For the finite motion minimum of the potential $U_+(\varrho)$ is reached at and is equal to 
\begin{equation}
\label{eq:quantum-potential-minimum}
\begin{aligned}
\varrho^*&=\left(\frac{m^2}{n}\right)^{1/(n+2)}, \\
\mathcal{E}^*&=U_+(\varrho^*)=\left(\frac{n}{2}+1\right)\left(\frac{m^2}{n}\right)^{n/(n+2)}.
\end{aligned}
\end{equation}
Assuming  $\mathcal{E}\gg \mathcal{E}^*$, we find the turning points (solutions of $U_+(\varrho)=\mathcal{E}$)
\begin{equation}
\label{eq:quantum-potential-turning}
\begin{gathered}
\varrho_1=\frac{\left|m\right|}{\sqrt{2\mathcal{E}}}\ll \varrho^*
\text{ and }  
\varrho_2=\mathcal{E}^{1/n}\gg \varrho^*,
\end{gathered}
\end{equation}
where we assumed that $\left|m\right|\ll\mathcal{E}^{(n+2)/2n}$.

At last, we find the wave function for the finite motion in $\{xy\}$ plane
\begin{equation}
\label{eq:final}
\psi(x,y)=F_m\left(\frac{\rho}{a_0}\right) \frac{e^{im\varphi}}{\sqrt{\rho}} \begin{pmatrix} e^{in\varphi/2} \\ ie^{-in\varphi/2}\end{pmatrix},
\end{equation}
where function $F_m(\rho/a_0)$ depends on energy $\varepsilon$.
Let's compare \eqref{eq:final} with spinor
\begin{equation}
\label{eq:spinor}
\Xi=e^{i\alpha} \begin{pmatrix} \cos(\theta/2) e^{-i\phi/2} \\ \sin(\theta/2) e^{i\phi/2}\end{pmatrix},
\end{equation}
corresponding to wave functions of spin-1/2 particle, directed along vector $\bm \zeta$, and $\alpha$ is an arbitrary phase. This vector is described by polar angles $\theta$ and $\phi$. We observe that
\begin{align}
\label{eq:angles}
\phi&=\frac{\pi}{2}-n\varphi, & \theta&=\frac{\pi}{2}.
\end{align}
Now, taking a scalar product of $\mathbf{B}$~\eqref{eq:field-0} and $\bm \zeta=\{\cos(\pi/2-n\varphi),\sin(\pi/2-n\varphi),0\}$ we obtain
\begin{equation}
\mathbf{B}\cdot\bm \zeta=G_n\rho^n=\left|\mathbf{B}\right|\left|\bm \zeta\right|.
\end{equation}
\\
For the infinite motion, the wave function is
\begin{equation}
\label{eq:final-inf}
\widetilde\psi(x,y)=\widetilde F_m\left(\frac{\rho}{a_0}\right) \frac{e^{im\varphi}}{\sqrt{\rho}} \begin{pmatrix} e^{in\varphi/2} \\ -ie^{-in\varphi/2}\end{pmatrix}.
\end{equation}
Comparison of \eqref{eq:final-inf} with spinor~\eqref{eq:spinor} gives
\begin{align}
\label{eq:angles-inf}
\phi&=-\frac{\pi}{2}-n\varphi, & \theta&=\frac{\pi}{2},
\end{align}
so that the scalar product of $\mathbf{B}$ and $\bm \zeta=\{\cos(-\pi/2-n\varphi),\sin(-\pi/2-n\varphi),0\}$ is
\begin{equation}
\mathbf{B}\cdot\bm \zeta=-G_n\rho^n=-\left|\mathbf{B}\right|\left|\bm \zeta\right|.
\end{equation}

Notice that expressions in \eqref{eq:angles} and \eqref{eq:angles-inf} are independent of quantum number $m$ and energy $\varepsilon$; vector $\bm \zeta$ for finite motion is parallel (magnetic moment is anti-parallel) to magnetic field $\bm B(\bm r)$ and anti-parallel $\bm B(\bm r)$ for infinite motion (magnetic moment is parallel).

Very large quantum numbers ($m$) and independence of spin direction on energy and $m$ allow transition from quantum description of the problem to classical. At first we need to construct a wave packet, a superposition of wave functions \eqref{eq:final} or \eqref{eq:final-inf} with different values of $\varepsilon$ and $m$, such that their spreads $\Delta\varepsilon\ll\langle\varepsilon\rangle$ and $\Delta m\ll\langle m\rangle$ are far less than the average values. Motion of the packet center corresponds to trajectory in the classical physics. The trajectory in the case of finite motion is described by the classical Hamiltonian
\begin{equation}
\label{eq:ham-fin}
H_{cl}^{f}=\frac{\mathbf{p}^2}{2M} + \left|\mu\right| \left|\boldsymbol{B}(\boldsymbol{r})\right|,
\end{equation}
and in the case of infinite motion
\begin{equation}
\label{eq:ham-infin}
H_{cl}^{inf}=\frac{\mathbf{p}^2}{2M} - \left|\mu\right| \left|\boldsymbol{B}(\boldsymbol{r})\right|,
\end{equation}
where we the relation $G_n\rho^n=\left|\boldsymbol{B}(\boldsymbol{r})\right|$ is used. We emphasize that Eqs.~\eqref{eq:ham-fin} and \eqref{eq:ham-infin} are valid not only for $2(n+1)$ pole magnet, but also for macroscopic magnetic fields of any configurations. Validity of such treatment for macroscopic fields is justified by infinitesimal variation of the magnetic field at the distance comparable with de Broglie wavelength. Thus, on each trajectory (finite and infinite) spin follows direction of the magnetic field. 
On the other hand, there is a well known equation for spin precession in a magnetic field
\begin{equation}
\label{eq:spin-precession}
\dot{\mathbf{S}}=\frac{2\mu}{\hbar}\left[\mathbf{S}\times\mathbf{B}\right],
\end{equation}
where direction of $\mathbf{S}$ is arbitrary with respect to $\mathbf{B}$. Resolution of apparent contradiction is the following.
In order to transit from quantum to classical mechanics we need to build a wave packet from the stationary solutions. 
Denoting the wave packets for finite and infinite motions with the same average momentum  $\mathbf{P}$ at the time $t_0$ as $\psi_f(\mathbf{r},t)$ and $\psi_{inf}(\mathbf{r},t)$ respectively,
we describe neutron by localized wave packet $\psi_0(\mathbf{r},t)$, which at time $t_0$ also has average momentum $\mathbf{P}$, and spin is non-collinear to the magnetic field. 
The centers of all packets coincide. 
Thus,  $\psi_0(\mathbf{r},t)=a \psi_f(\mathbf{r},t) + b \psi_{inf}(\mathbf{r},t)$, where coefficients $a$ and $b$ depend on spin orientation at time $t_0$. The average spin value $\left<\mathbf{s}(t)\right>$ at time $t>t_0$ is 
\begin{equation}
\label{eq:spin-interference}
\begin{split}
\left<\mathbf{s}(t)\right>&=\left< a \psi_f(\mathbf{r},t) + b \psi_{inf}(\mathbf{r},t) \right| \mathbf{s} \\
&\mspace{162mu} \left| a \psi_f(\mathbf{r},t) + b \psi_{inf}(\mathbf{r},t) \right> \\
&=\left|a\right|^2\left< \psi_f(\mathbf{r},t) \right| \mathbf{s} \left| \psi_f(\mathbf{r},t) \right> \\
&\quad+ \left|b\right|^2\left< \psi_{inf}(\mathbf{r},t) \right| \mathbf{s} \left| \psi_{inf}(\mathbf{r},t) \right> \\
&\quad+2Re
\left[ 
a^*b\left< \psi_f(\mathbf{r},t) \right| \mathbf{s} \left| \psi_{inf}(\mathbf{r},t) \right>
\right].
\end{split}
\end{equation}
If during time $\delta t=t-t_0$ the centers of the packets $\psi_f(\mathbf{r},t)$ and $\psi_{inf}(\mathbf{r},t)$ diverge by distance $\delta r$ much greater than size $\sigma_{wp}$  of the packet $\psi_0(\mathbf{r},t_0)$ at time $t_0$, then the interference term disappears and we are left with two classical trajectories, finite and infinite, having corresponding spins (parallel and anti-parallel to $\mathbf{B}$). Otherwise, if during $\delta t$ the distance $\delta r \lesssim \sigma_{wp}$, then interference is important and we are having one wave packet with precessing average spin. 

The size of the wave packet could be estimated as the size of the quantim oscillator coherent state \cite{Landau:1977} with oscillation frequency (paragraph~\ref{sct:miltipole})
\begin{equation}
\omega\sim\sqrt{\frac{ \left|\mu\right| B_0}{M\rho_0^2}},
\end{equation}
\begin{equation}
\label{eq:wp-size}
\sigma_{wp}=\left(\frac{\hbar}{2M\omega}\right)^{1/2}\sim\left( \frac{\hbar^2 \rho_0^2}{M\left|\mu\right| B_0} \right)^{1/4},
\end{equation}
where $B_0$ is characteristic magnetic field at radius $\rho_0$. Note that for sextupole $G_2=B_0/\rho_0^2$ and $\rho_0$ disappears from $\sigma_{wp}$ in \eqref{eq:wp-size}. The distance between diverging trajectories is
\begin{equation}
\label{eq:div-distance}
\delta r \sim \frac{\left|\mu\right| B_0}{M \rho_0} {\delta t}^2=\frac{\left|\mu\right| B_0}{E \rho_0}L^2,
\end{equation}
where $\delta t\approx L M/p$, $p=\sqrt{2M E}$, $L$ is longitudinal size of the magnet, and longitudinal speed is much greater than transverse one.
Hence, application condition of our solution and existence of SGE ( $\delta r \gtrsim \sigma_{wp}$) is 
\begin{equation}
\label{eq:apl-condition}
L\gtrsim \left( \frac{\hbar^2 E^4}{M} \frac{\rho_0^6}{\left|\mu\right|^5 B_0^5} \right)^{1/8},
\end{equation}
which requires low energies and high gradients $B_0/\rho_0$. For the case described in section~\ref{sct:Benchmark} with  $E=0.033$~eV, $G=400$~T/m, $\rho_0=2$~mm, the condition yields $L\gtrsim 10$~cm.

Let us recall, that equations \eqref{eq:ham-fin} and \eqref{eq:ham-infin} are valid only for neutral particles, since for charged particles (electron or proton) in macroscopic fields the spin part of the magnetic moment is insignificant with respect to the total magnetic moment. As a result, there is no unpolarized beam splitting into two polarized \cite{Kessler:1985}.

Advantage of our approach, using Hamiltonians \eqref{eq:ham-fin} and \eqref{eq:ham-infin}, is that we reduce the motion of neutral spin-1/2 particle in the magnetic field to the classical motion of spinless particle in the potential well.

Since momentum component $p_z$ is conserved, and momentum direction in $\{xy\}$ plane is changed, then the total classical momentum $\boldsymbol{p}$ will also change its direction. This observation opens a possibility for storage ring creation by using a combination of lenses.

In the following sections we compare numerical results of our approach with reported results of neutron experiments \cite{Sherwood:1954,PTA1965en} and present conceptual design of a storage ring based on Hamiltonians \eqref{eq:ham-fin} and \eqref{eq:ham-infin}.

\section{Comparison with neutron beam splitting experiment}
\label{sct:Benchmark}
The authors of \cite{PTA1965en} conducted Stern-Gerlach experiment with neutron beam and published results together with the detailed description of the experimental setup, which allows us to verify our approach. 
The authors reported neutron beam splitting in the field $B_y=B_0+G_1y$ with $B_0=-0.8$~T, $G_1=400$~T/m. The magnet length was $L=0.5$~m. The beam parameters were: energy $E=0.033$~eV, energy spread is not reported (we assume none), horizontal width $\Delta x=2.5$~mm, vertical width $\Delta y=0.25$~mm (we assume uniform distribution), horizontal angular spread $\alpha_x=\pm0.4'=\pm1.16\times 10^{-4}$, vertical $\alpha_y=\pm1.16\times 10^{-4}$ (we assume normal distribution with 3 standard deviations). Observed beam deflection was $\pm 2.3\times 10^{-4}$. The approximate number of detected particles was $10^4$.

\begin{figure}[htbp]
\centering
\includegraphics*[width=\columnwidth,trim=0 0 0 0, clip]{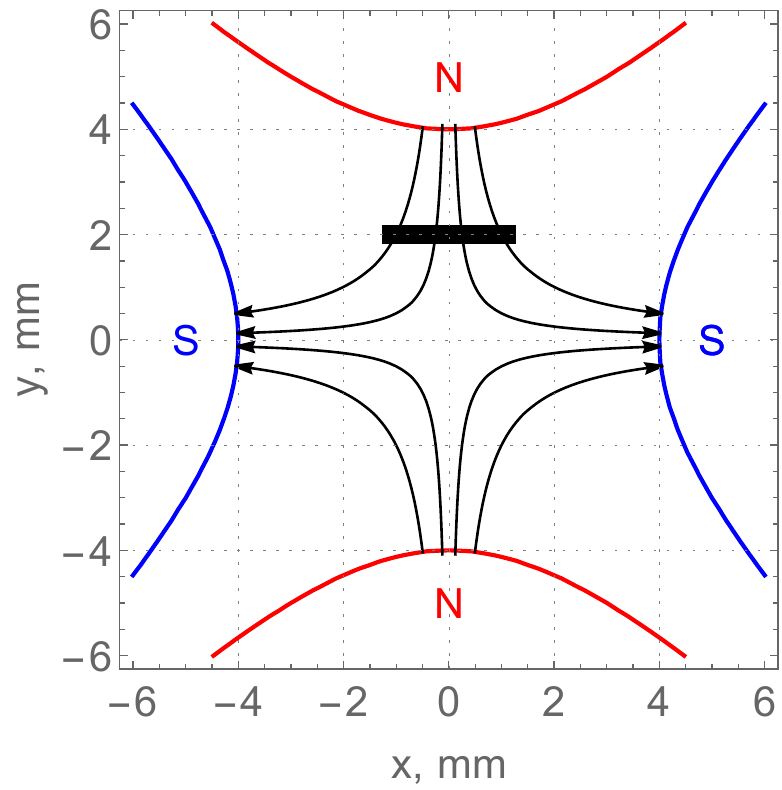}
\caption{The scheme of simulated magnet (skew quadrupole), magnetic field lines, position, shape and dimensions of the beam (the black rectangle).}
\label{fig:skewquad-scheme}
\end{figure}

In order to test our solution, we performed calculation of magnetic field in the magnet \cite{PTA1965en} with the help of COMSOL Multiphysics\textsuperscript{\textregistered} software~\cite{comsol}. The magnet field was found to be significantly nonlinear. Since the authors reported only the values of the constant field and its gradient, we represent the original magnet as a skew quadrupole with identical $G_1=400$~T/m, while the beam is shifted vertically with respect to symmetry plane by $2$~mm providing similar $B_0=-0.8$~T.
The field of such a magnet is
\begin{equation}
\label{eq:skewquad-field}
\begin{gathered}
B_x=G_1x, \mspace{18mu} B_y=-G_1y, \mspace{18mu} B_z=0.
\end{gathered}
\end{equation}
FIG.~\ref{fig:skewquad-scheme} shows the scheme of the simulated magnet and neutron beam position and dimensions. The beam propagates along the $z$ axis.

Substitution of the fields \eqref{eq:skewquad-field} into \eqref{eq:ham-fin} and \eqref{eq:ham-infin} yields Hamiltionians for skew quadrupole,
\begin{align}
\label{eq:ham-fin-skew}
H_{sq}^{f}&=\frac{\mathbf{p}^2}{2M} + \left|\mu\right| \left|G_1\right|\sqrt{x^2+y^2}, \\
\label{eq:ham-infin-skew}
H_{sq}^{inf}&=\frac{\mathbf{p}^2}{2M} - \left|\mu\right| \left|G_1\right|\sqrt{x^2+y^2}.
\end{align}
The equations of motions are
\begin{equation}
\left\{
\begin{aligned}
\dot{z}&=\frac{p_z}{M} &\Rightarrow && z&=z_0+\frac{p_{z,0}}{M}t,\\
\dot{p_z}&=0 &\Rightarrow && p_z&=p_{z,0},
\end{aligned}
\right.
\label{eq:z-motion-1}
\end{equation}
\begin{equation}
\left\{
\begin{aligned}
\dot{x}&    =\frac{p_x}{M} \\
\dot{p}_x&=\mp \frac{ \left|\mu\right| \left|G_1\right|x}{\sqrt{x^2+y^2}},
\end{aligned}
\right.
\label{eq:x-motion-1}
\end{equation}
\begin{equation}
\left\{
\begin{aligned}
\dot{y}&    =\frac{p_y}{M} \\
\dot{p}_y&=\mp\frac{ \left|\mu\right| \left|G_1\right|y}{\sqrt{x^2+y^2}},
\end{aligned}
\right.
\label{eq:y-motion-1}
\end{equation}
where $p_{z,0}=\sqrt{2ME}$ is the initial condition, and the top sign in $p_{x,y}$ equations describes the case of the finite motion.
The initial conditions for numerical simulations were chosen as: 
uniform distribution with given width and expected values $\langle x\rangle=0$, $\langle y\rangle=2$~mm for coordinates $\{x,y\}$;
normal distribution with $\sigma=\alpha_{x,y} p_{z,0}/3$ and expected values $\langle p_{x,y}\rangle=0$ for momenta $\{p_x,p_y\}$.

Note that equations \eqref{eq:x-motion-1} and \eqref{eq:y-motion-1} differ from equations (3) and (4) reported in \cite{Lee:2024wgh} and there is no uniform force.

Equations \eqref{eq:x-motion-1} and \eqref{eq:y-motion-1} for horizontal and vertical motion were solved numerically with the help of Wolfram Mathematica \cite{Mathematica}.

\begin{figure}[htbp]
\centering
\includegraphics*[width=\columnwidth,trim=0 0 0 0, clip]{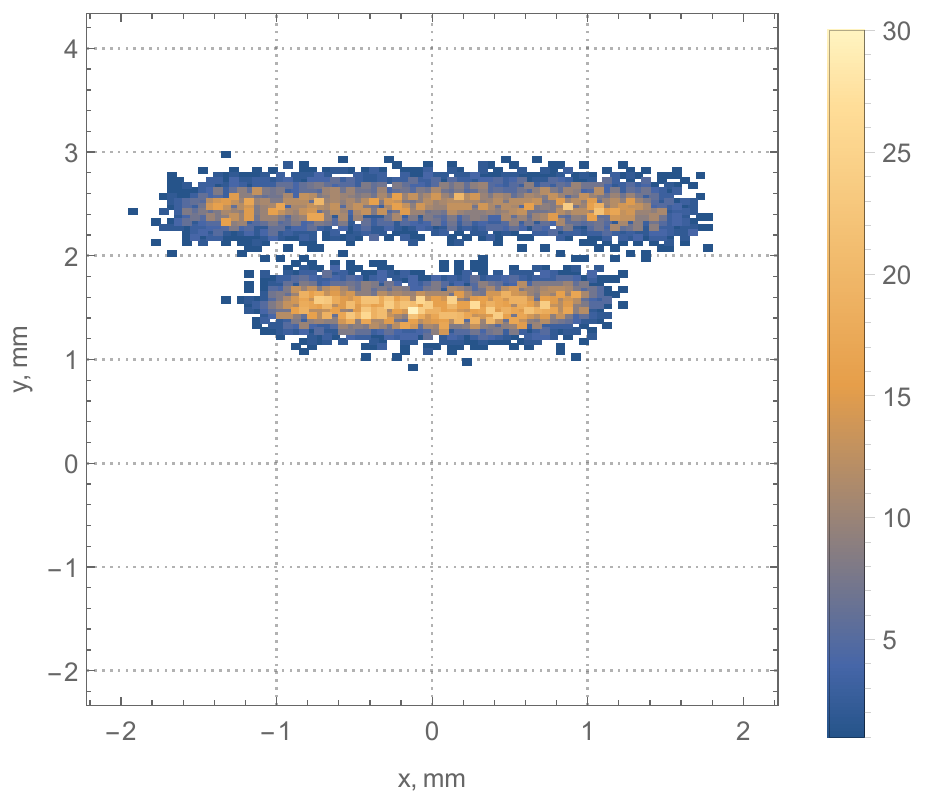}
\caption{2d coordinate beam distribution in the detector plane at distance of 256~cm. The top (larger) spot corresponds to Hamiltonian \eqref{eq:ham-infin-skew}, the bottoms (smaller) corresponds to Hamiltonian \eqref{eq:ham-fin-skew}. The color denotes the number of particles.}
\label{fig:skewquad-coordhist-end-xy}
\end{figure}

\begin{figure}[htbp]
\centering
\includegraphics*[width=\columnwidth,trim=0 0 0 0, clip]{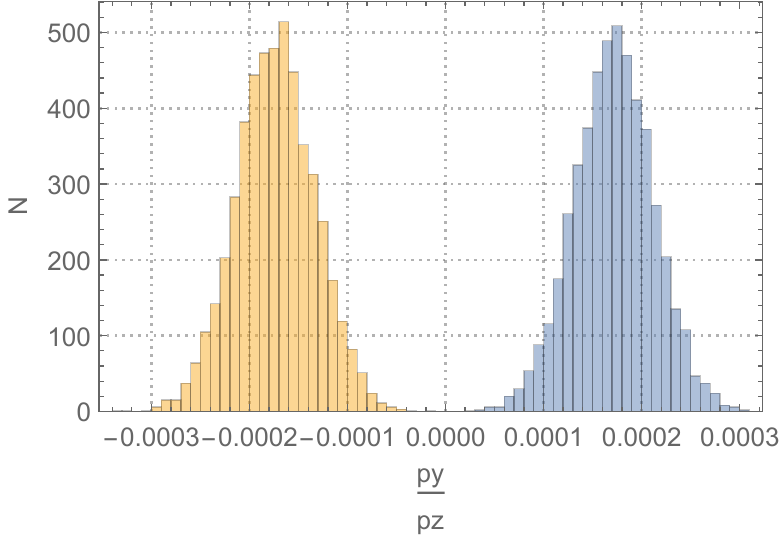}
\caption{Normalized vertical momentum beam distribution behind the skew quadrupole.}
\label{fig:skewquad-momentumhist-end-py}
\end{figure}

One half of the initial beam (black rectangle on FIG.~\ref{fig:skewquad-scheme}) was propagated through the skew quadrupole. Its trajectory corresponds to the Hamiltonian \eqref{eq:ham-fin-skew} describing particles with spin directed along the field. The motion of the other half is described by the Hamiltonian \eqref{eq:ham-infin-skew} corresponding to spin opposite to the field direction. 
FIG.~\ref{fig:skewquad-coordhist-end-xy} and FIG.~\ref{fig:skewquad-momentumhist-end-py} show resulting coordinate distribution in the detector plane and normalized momentum distributions, respectively, after the skew quadrupole. 

The average deflection angle obtained in our simulation is $\pm(1.7\pm0.4)\times 10^{-4}$ which is close to $\pm 2.3\times 10^{-4}$ reported in \cite{PTA1965en}. The difference of about $30$\% in deflection angles could be explained by presence of higher multipoles in the original magnet \cite{PTA1965en} and by uncertainties of the beam initial conditions.

\section{Neutron motion in multipole magnet}
\label{sct:miltipole}
Dipole magnet does not affect neutron motion. In charged particle accelerators quadrupole and sextupole magnets are routinely used, and advanced technologies of magnet manufacturing make it possible to achieve high gradients necessary to control neutron  motion. Therefore, below we consider neutron motion in quadrupole and sextupole magnets.
\subsection{Neutron trajectories in quadrupole}
Hamiltonians and corresponding equations of motion for neutron in the quadrupole are identical to those for the skew quadrupole \eqref{eq:ham-fin-skew} and \eqref{eq:ham-infin-skew}, \eqref{eq:z-motion-1}, \eqref{eq:x-motion-1} and \eqref{eq:y-motion-1}. Since, there is no simple analytic solution in Cartesian coordinate system (transverse motion is intrinsically coupled), we use cylindrical coordinate system $\{\rho,\varphi,z\}$, in which Hamiltonian reads
\begin{equation}
\label{eq:ham-cyl}
H_q=\frac{p_z^2}{2M}+\frac{p_\rho^2}{2M}+\frac{p_\varphi^2}{2M\rho^2}\pm\left|\mu\right| \left|G_1\right|\rho,
\end{equation}
where $x=\rho \cos\varphi$, $y=\rho \sin\varphi$, $p_\rho=p_x\cos\varphi+p_y\sin\varphi$, $p_\varphi=p_y x- p_x y$, the top sign ``$+$'' corresponds to finite motion, the bottom ``$-$'' to infinite. The angular momentum is time independent $p_\varphi=const$ because there is no direct dependance on $\varphi$ in \eqref{eq:ham-cyl}, but the angular velocity obeys equation
\begin{equation}
\dot{\varphi}=\frac{p_\varphi}{M\rho^2}.
\end{equation}
The finite motion is governed by the potential
\begin{equation}
\label{eq:potential-fin}
U(\rho)=\frac{p_\varphi^2}{2M\rho^2}+\left|\mu\right| \left|G_1\right|\rho,
\end{equation}
which minimum $\rho^*$ and turning points $\rho_{1,2}$ are found in \eqref{eq:quantum-potential-minimum} and \eqref{eq:quantum-potential-turning}, where angular momentum is $p_\varphi= m \hbar$ ($m\gg1$). 
The radial equations of motion are
\begin{equation}
\label{eq:quad-rho-fin}
\left\{
\begin{aligned}
\dot{\rho}&=\frac{p_\rho}{M} \\
\dot{p_\rho}&=\frac{p_\varphi^2}{M\rho^3}-\left|\mu\right| \left|G_1\right|.
\end{aligned}
\right.
\end{equation}
Introducing
\begin{equation}
\label{eq:quad-E-perp}
E_\perp=\frac{p_{\rho,0}^2}{2M}+\frac{p_\varphi^2}{2M\rho_0^2}+\left|\mu\right| \left|G_1\right|\rho_0=const,
\end{equation}
with $\rho_0$ being initial condition, we find: turning points $\rho_{1,2}$ as solutions of
\begin{equation}
\label{eq:quad-turn-points}
E_\perp=\frac{p_\varphi^2}{2M\rho^2}+\left|\mu\right| \left|G_1\right|\rho,
\end{equation}
and period of radial oscillations according to
\begin{equation}
\label{eq:quad-period}
T_\rho=\sqrt{2M}\int_{\rho_1}^{\rho_2}\frac{d\rho}{\sqrt{E_\perp-\dfrac{p_\varphi^2}{2M\rho^2}-\left|\mu\right| \left|G_1\right|\rho}}.
\end{equation}
The oscillation period, corresponding frequency $\omega_\rho=2\pi/T_\rho$ and spatial period $\lambda_\rho=v_z T_\rho$ have a strong dependence on initial conditions (property of nonlinear oscillations) and have no simple analytic representation.

\begin{figure}[htbp]
\centering
\includegraphics*[width=\columnwidth,trim=0 0 0 0, clip]{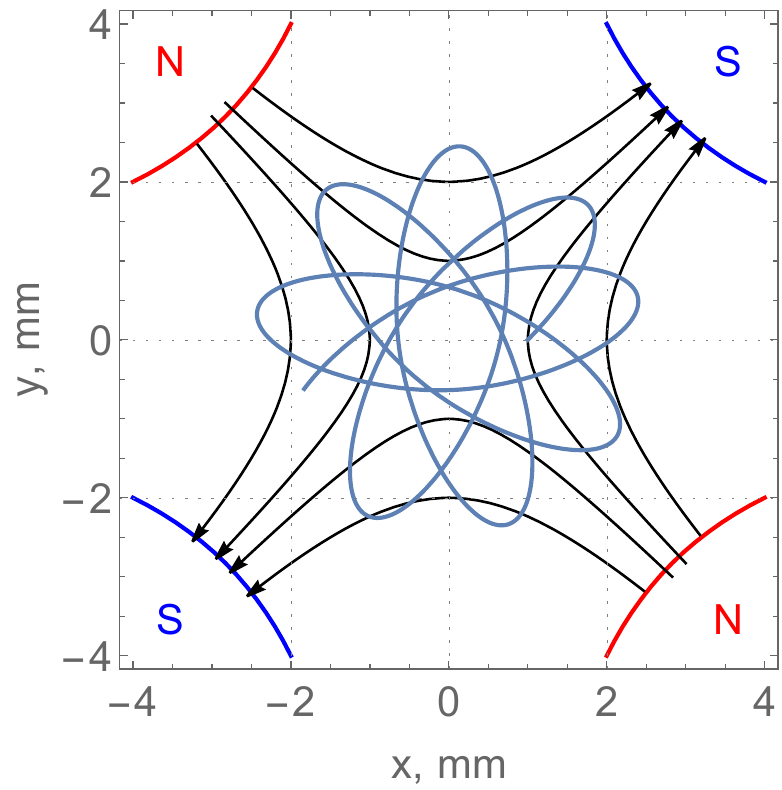}
\caption{An example of neutron trajectory (dark blue) in the quadrupole with $G=100$~T/m, $E=10^{-7}$~eV, $x_0=1$~mm, $y_0=0$~mm,  $p_{x,0}c=3$~eV, $p_{y,0}c=3$~eV, quadrupole poles are in blue and red, field lines are in black.}
\label{fig:quad-traj-xy-1}
\end{figure}

For the quadrupole with gradient $G_1=100$~T/m and neutron with energy $E=10^{-7}$~eV and initial conditions $x_0=1$~mm, $y_0=0$~mm, $p_z c=13.71$~eV ($v_z=4.37$~m/s), $p_{x,0}c=3$~eV, $p_{y,0}c=3$~eV numerically found trajectory is shown in FIG.~\ref{fig:quad-traj-xy-1}. Note that trajectory never closes, and the spin direction is always along the field lines on the particle's trajectory.

The infinite motion is governed by the potential
\begin{equation}
\label{eq:potential-infin}
U(\rho)=\frac{p_\varphi^2}{2M\rho^2}-\left|\mu\right| \left|G_1\right|\rho,
\end{equation}
which does not form a potential well. 

Since in both cases of finite or infinite neutron motion in the quadrupole, trajectory is not an arc of the circle as it is for electron in the bending magnet, its use for storage ring would be difficult.

\begin{figure}[htbp]
\centering
\includegraphics*[width=\columnwidth,trim=0 0 0 0, clip]{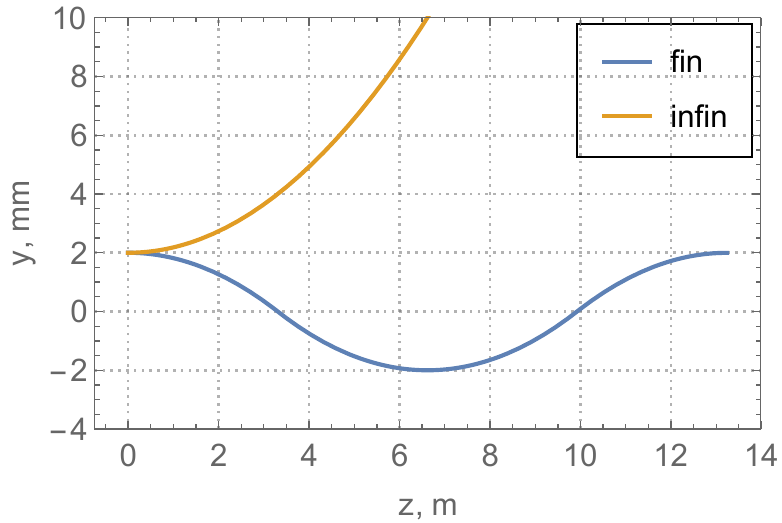}
\caption{Beam vertical trajectories in the skew quadrupole as a function of the magnet length.}
\label{fig:skewquad-y-traj-fin-infin-6.52-3}
\end{figure}

Now let's consider what was observed in Stern-Gerlach experiment \cite{PTA1965en} (section~\ref{sct:Benchmark}), the beam splitting in the vertical direction was symmetrical with respect to the beam entrance position in the magnet (FIG.~\ref{fig:skewquad-coordhist-end-xy} and FIG.~\ref{fig:skewquad-momentumhist-end-py}), because the length of the magnet $L=0.5$~m was smaller than $\lambda_\rho\approx 6.62$~m.
For the magnet length $L=6.62$~m the deflection angle of the infinite trajectory is significantly larger $(2.3\pm0.13)\times 10^{-3}$ than for the finite one $(-6.6\pm7.8)\times 10^{-5}$. The finite and infinite trajectories inside the magnet as a function of magnet length are shown in  FIG.~\ref{fig:skewquad-y-traj-fin-infin-6.52-3}.

\subsection{Neutron trajectories in sextupole}
Sextupole field and Hamiltonians are
\begin{equation}
\label{eq:sextupole-field-ham}
\begin{gathered}
B_x=Sxy, \mspace{18mu} B_y=S\frac{\left(x^2-y^2\right)}{2}, \mspace{18mu} B_z=0, \\
H_{sext}=\frac{\mathbf{p}^2}{2M} \pm \left|\mu\right| \left|S\right|\frac{x^2+y^2}{2},
\end{gathered}
\end{equation}
where the top sign ``$+$'' and the bottom ``$-$'' describe finite and infinite motion, respectively, $S=2G_2$.
Equations of longitudinal motion are identical to \eqref{eq:z-motion-1}, equations of transverse motion are
\begin{equation}
\left\{
\begin{aligned}
\dot{x}&    =\frac{p_x}{M} \\
\dot{p}_x&=\mp  \left|\mu\right| \left|S\right|x,
\end{aligned}
\right.
\label{eq:sext-x-motion}
\end{equation}
\begin{equation}
\left\{
\begin{aligned}
\dot{y}&    =\frac{p_y}{M} \\
\dot{p}_y&=\mp \left|\mu\right| \left|S\right|y.
\end{aligned}
\right.
\label{eq:sext-y-motion}
\end{equation}
These equations have simple uncoupled solutions:
for finite motion
\begin{equation}
\label{eq:sext-sol-fin}
\left\{
\begin{aligned}
x&=x_0\cos(\omega t)+\frac{p_{x,0}}{M\omega}\sin(\omega t) \\
p_x&=-x_0M\omega\sin(\omega t)+p_{x,0}\cos(\omega t) \\
y&=y_0\cos(\omega t)+\frac{p_{y,0}}{M\omega}\sin(\omega t) \\
p_y&=-y_0M\omega\sin(\omega t)+p_{y,0}\cos(\omega t),
\end{aligned}
\right.
\end{equation}
for infinite motion
\begin{equation}
\label{eq:sext-sol-inf}
\left\{
\begin{aligned}
x&=x_0\cosh(\omega t)+\frac{p_{x,0}}{M\omega}\sinh(\omega t) \\
p_x&=x_0M\omega\sinh(\omega t)+p_{x,0}\cosh(\omega t) \\
y&=y_0\cosh(\omega t)+\frac{p_{y,0}}{M\omega}\sinh(\omega t) \\
p_y&=y_0M\omega\sinh(\omega t)+p_{y,0}\cosh(\omega t),
\end{aligned}
\right.
\end{equation}
where subscript $0$ denotes initial conditions,
\begin{equation}
\label{eq:sext-omega}
\omega=\sqrt{\frac{\left|\mu\right| \left|S\right|}{M}}.
\end{equation}
Note that trajectories in cylindrical coordinates are ellipses for finite motion and hyperbolas for infinite motion. The center of ellipse coincides with the symmetry axis of the sextupole; therefore, sextupole plays the role of the focusing (defocusing) lens but not the bending magnet.

\begin{figure}[htbp]
\centering
\includegraphics*[width=\columnwidth,trim=0 0 0 0, clip]{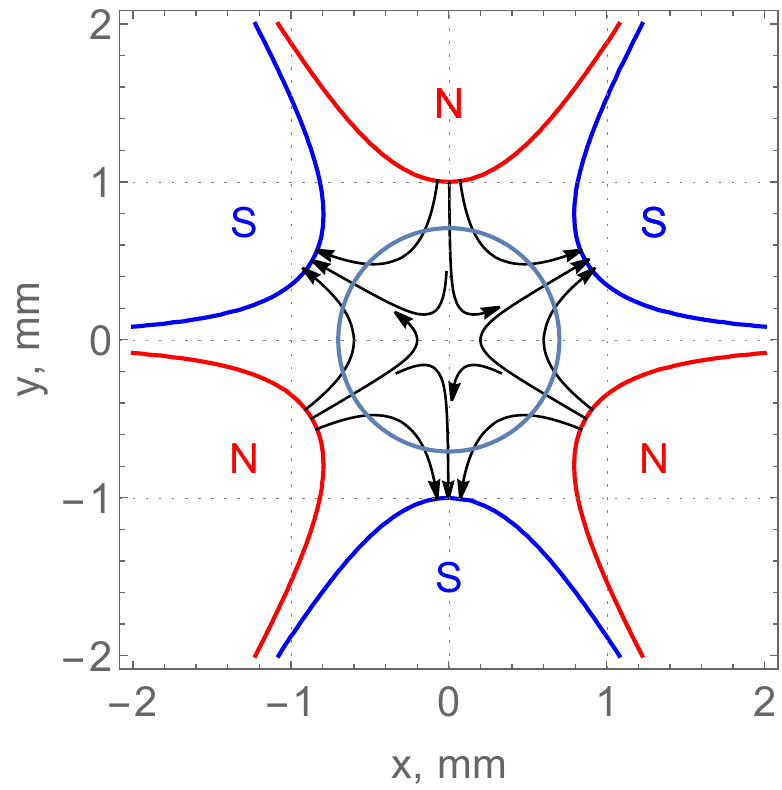}
\caption{An example of neutron trajectory (dark blue) in the sextupole with $S=4\times 10^5\text{ T/m}^2$, $E=10^{-3}$~eV, $x_0=0.7$~mm, $y_0=0$~mm,   $p_{x,0}c=0$~eV, $p_{y,0}c=3.37$~eV ($v_{y,0}=1.07$~m/s), sextupole poles are in blue and red, field lines are in black. Spin direction is always along the field lines on the particle's trajectory.}
\label{fig:sext-traj-xy-1}
\end{figure}

\begin{figure}[htbp]
\centering
\includegraphics*[width=\columnwidth,trim=0 0 0 0, clip]{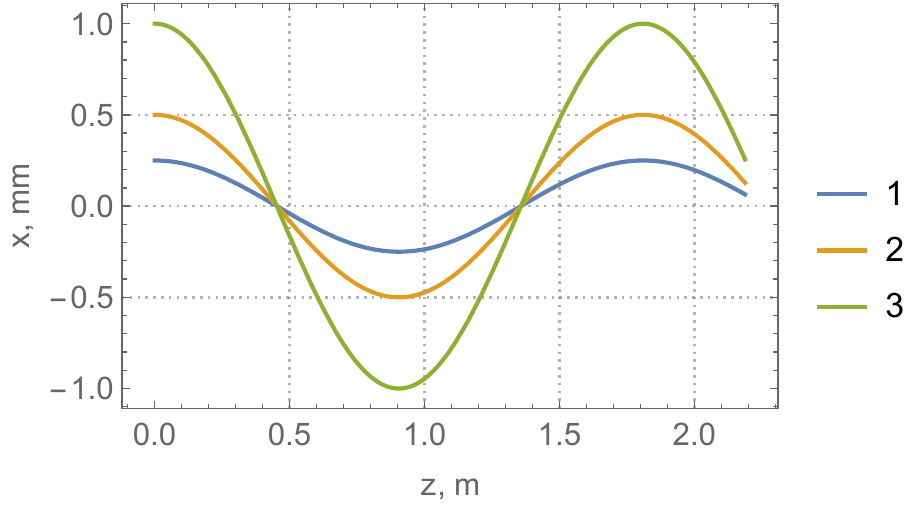}
\caption{Neutron horizontal trajectory (finite) dependence on traveled distance $z$ in the sextupole with $S=4\times 10^5\text{ T/m}^2$, $E=10^{-3}$~eV, $x_0=\{0.25,0.5,1\}$~mm (blue,yellow,green), $y_0=0$~mm,  $p_{x,0}c=0$~eV, $p_{y,0}c=0$~eV.}
\label{fig:sext-traj-x-z-1}
\end{figure}

\begin{figure}[htbp]
\centering
\includegraphics*[width=\columnwidth,trim=0 0 0 0, clip]{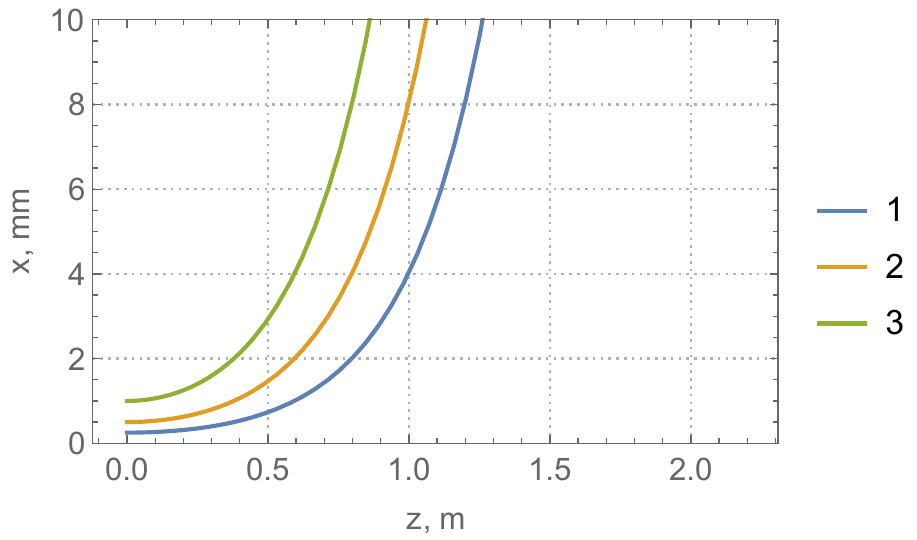}
\caption{Neutron horizontal trajectory (infinite) dependence on traveled distance $z$ in the sextupole with $S=4\times 10^5\text{ T/m}^2$, $E=10^{-3}$~eV, $x_0=\{0.25,0.5,1\}$~mm (blue,yellow,green), $y_0=0$~mm,  $p_{x,0}c=0$~eV, $p_{y,0}c=0$~eV.}
\label{fig:sext-traj-inf-x-z-1}
\end{figure}

The traveled distance in the magnet relates to the period of transverse oscillations as
\begin{equation}
\label{eq:sext-lambda}
\lambda=v_z\frac{2\pi}{\omega}=\frac{p_{z,0}}{M}2\pi \sqrt{\frac{M}{\left|\mu\right|\left|S\right|}}.
\end{equation}
FIG.~\ref{fig:sext-traj-xy-1} shows neutron trajectory and field lines in the sextupole for particular initial conditions $E=10^{-3}$~eV, $x_0=0.7$~mm, $y_0=0$~mm,  $p_{x,0}c=0$~eV, $p_{y,0}c=3.37$~eV ($v_{y,0}=1.07$~m/s) and $S=4\times 10^5\text{ T/m}^2$.
Trajectories of the three particles in the sextupole with $S=4\times 10^5\text{ T/m}^2$, $E=10^{-3}$~eV, $x_0=\{0.25,0.5,1\}$~mm, $y_0=0$~mm,  $p_{x,0}c=0$~eV, $p_{y,0}c=0$~eV are shown in FIG.~\ref{fig:sext-traj-x-z-1} and in FIG.~\ref{fig:sext-traj-inf-x-z-1} for finite and infinite motions, respectively.

\subsection{Transport maps}
In order to design a storage ring, it is necessary to define a transformation of neutron coordinates from point $1$ to point $2$. Since equations \eqref{eq:sext-sol-fin} and \eqref{eq:sext-sol-inf} are linear, such a transformation is a transport matrix.
Introducing sextupole strength $K_1$, expressing time of flight $t$ through sextupole length $L$
\begin{equation}
\label{eq:sextupole-definitions}
\begin{aligned}
K_1&=\frac{\omega M}{p_z}, & t&=L\frac{M}{p_z}, & \omega t&=K_1L,
\end{aligned}
\end{equation}
using normalized transverse momenta ($p_z=const$)
\begin{equation}
\begin{aligned}
x'&=p_x/p_z, &  y'&=p_y/p_z,
\end{aligned}
\end{equation}
and vectors
\begin{equation}
\begin{aligned}
\mathbf{X}&=
\begin{pmatrix}
x \\
x'
\end{pmatrix},
&
\mathbf{Y}&=
\begin{pmatrix}
y \\
y'
\end{pmatrix}
\end{aligned}
\end{equation}
we can write sextupole transport matrix $\mathbf{R}$ via the relation $\mathbf{X_2}=\mathbf{R}\mathbf{X_1}$ (the same for $\mathbf{Y}$), as
\begin{equation}
\label{eq:sext-map-fin}
\mathbf{R}_{x,y}^{fin}=
\begin{pmatrix}
\cos(K_1L)     & \dfrac{\sin(K_1L)}{K_1} \\
-K_1\sin(K_1L) & \cos(K_1L)
\end{pmatrix}
\end{equation}
for finite motion, and
\begin{equation}
\label{eq:sext-map-inf}
\mathbf{R}_{x,y}^{inf}=
\begin{pmatrix}
\cosh(K_1L)     & \dfrac{\sinh(K_1L)}{K_1} \\
K_1\sinh(K_1L) & \cosh(K_1L)
\end{pmatrix}
\end{equation}
for infinite motion.
The matrices \eqref{eq:sext-map-fin} and \eqref{eq:sext-map-inf} describe focusing and defocusing elements, similar to charged particle optics (electron in quadrupole) \cite{HandbookAccelerator}. \\
The transport matrix of the drift of length $L$ is obvious,
\begin{equation}
\label{eq:sext-map-infin}
\mathbf{R}_{x,y}=
\begin{pmatrix}
1 & L \\
0 & 1
\end{pmatrix}.
\end{equation}
The transport matrix for the quadrupole is significantly more complicated due to the absence of simple analytical solution of equations of motion. Therefore, below we do not consider quadrupoles. 

\section{Proposal of testing experiment}
Neutron beam focusing with sextupole was performed in \cite{Yamada:2008zza}. Despite of successful demonstration, the sextupole was short, and splitting of finite and infinite trajectories was symmetrical. In order to observe the asymmetry between two cases, we propose to conduct an experiment at Budker INP.
For this we plan to use an existing accelerator-based neutron source VITA~\cite{biology10050350} developed for Boron Neutron Capture Therapy (BNCT).

\begin{figure*}
\includegraphics[width=\textwidth,trim=0 0 0 0, clip]{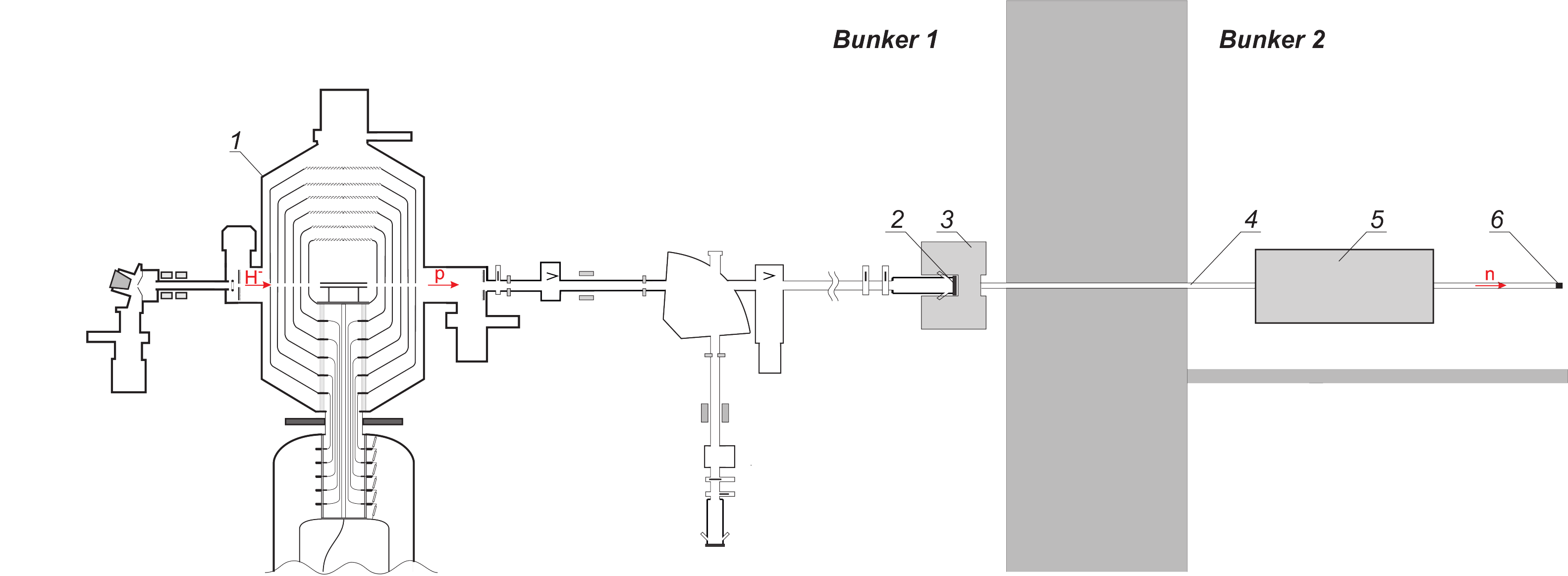}
\caption{\label{fig:vita} Layout of the experimental facility:  1 -- vacuum insulated tandem accelerator (VITA~\cite{biology10050350}), 2 -- lithium target, 3 -- moderator, 4 -- neutron guide, 5 -- sextupole magnet, 6 -- detector.}
\end{figure*}

Layout of experimental setup is shown in FIG.~\ref{fig:vita}.
The DC vacuum insulated tandem accelerator delivers 10~mA proton or deutron beam with energy 2.3~MeV to the lithium target, producing neutrons in $^7\text{Li(p,n)}^7\text{Be}$ or Li(d,n) reactions. The former yields neutrons with average energy of $0.2$~MeV at rate of $5\times 10^{12}\text{ s}^{-1}$, the latter produces neutrons with average energy of $6$~MeV at rate of $10^{13}\text{ s}^{-1}$. Neutrons are slowed down in moderator, and through neutron guide cold neutrons are delivered to the bunker, where a magnet and detector will be installed.
The spatial distribution of neutron beam is measured either by a neutron detector with a lithium or boron scintillator, or by a HPGe $\gamma$-spectrometer with samarium, cadmium or boron converter.

In order to distinguish trajectories of finite and infinite cases (FIG.~\ref{fig:sext-traj-x-z-1} and FIG.~\ref{fig:sext-traj-inf-x-z-1}) the beam should perform at least one spatial oscillation. The spatial period of the oscillation \eqref{eq:sext-lambda} depends on neutron beam energy, sextupole length and gradient. Choosing the magnet length equal to one spatial period at maximum gradient allows by varying the sextupole strength to observe oscillation of finite trajectory and divergence of the infinite. The spatial oscillation period \eqref{eq:sext-lambda} could be written as 
\begin{equation}
\lambda[\text{m}]=36183.7\sqrt{\frac{E[\text{ eV}]}{S[\text{ T/}\text{m}^2]}}.
\label{eq:sext-lambda-2}
\end{equation}
It follows from \eqref{eq:sext-lambda-2} that colder neutrons permit relaxed sextupole parameters (gradient, length, aperture). However, obtaining ultra-cold neutrons requires more effort: cryogenic technique, gravitation, etc. Thus, the choice of experimental setup is a compromise between neutron beam energy and magnet technologies.

\begin{figure}[htbp]
\centering
\includegraphics*[width=\columnwidth,trim=0 0 80 0, clip]{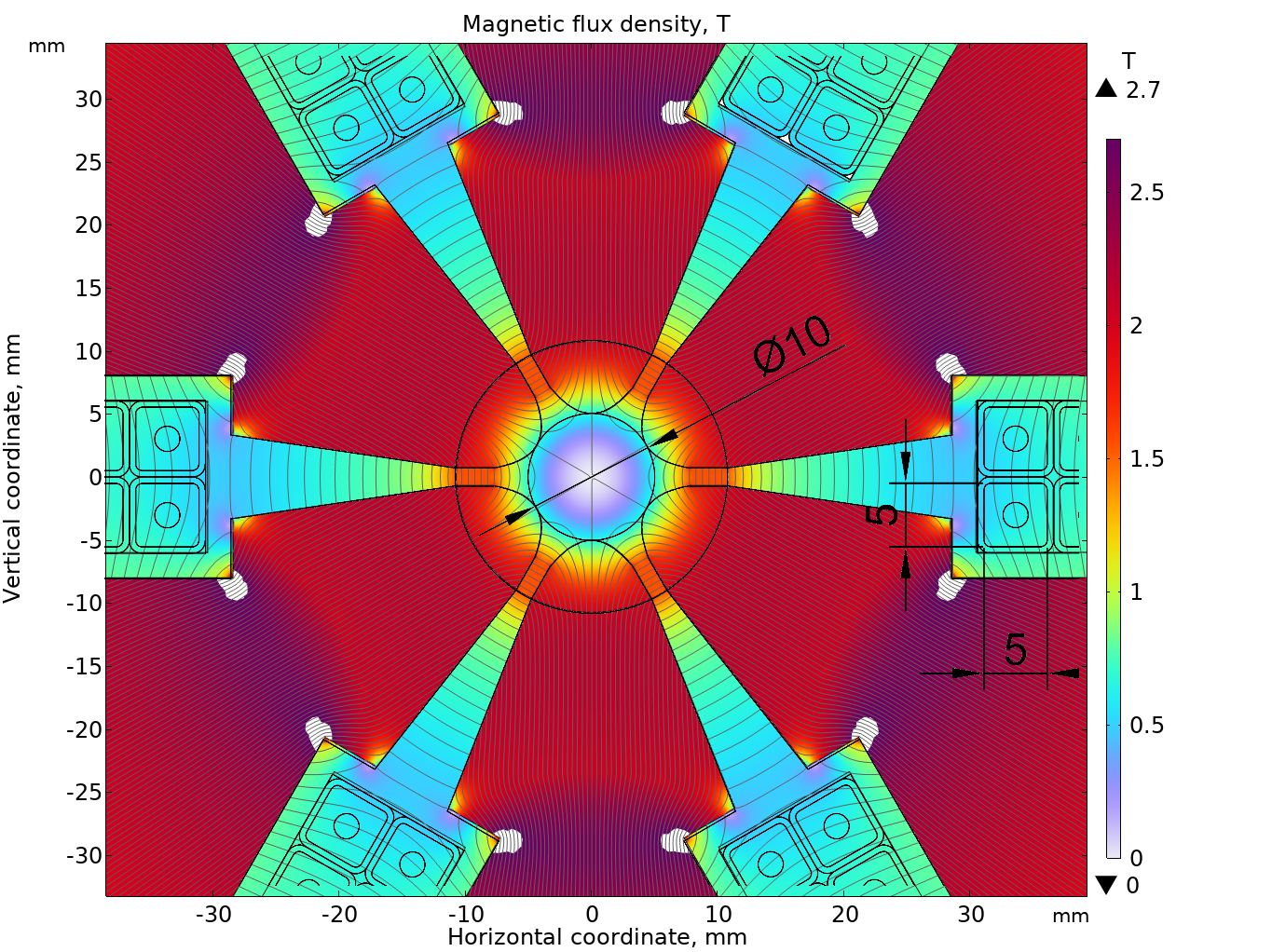}
\caption{Magnetic field (in color) of the proposed sextupole magnet with internal aperture of $R=5$~mm and $S=52\times 10^3\text{ T/m}^2$, close up of the working aperture.}
\label{fig:Proposal-sextupole-2}
\end{figure}

\begin{figure}[htbp]
\centering
\includegraphics*[width=\columnwidth,trim=120 15 570 15, clip]{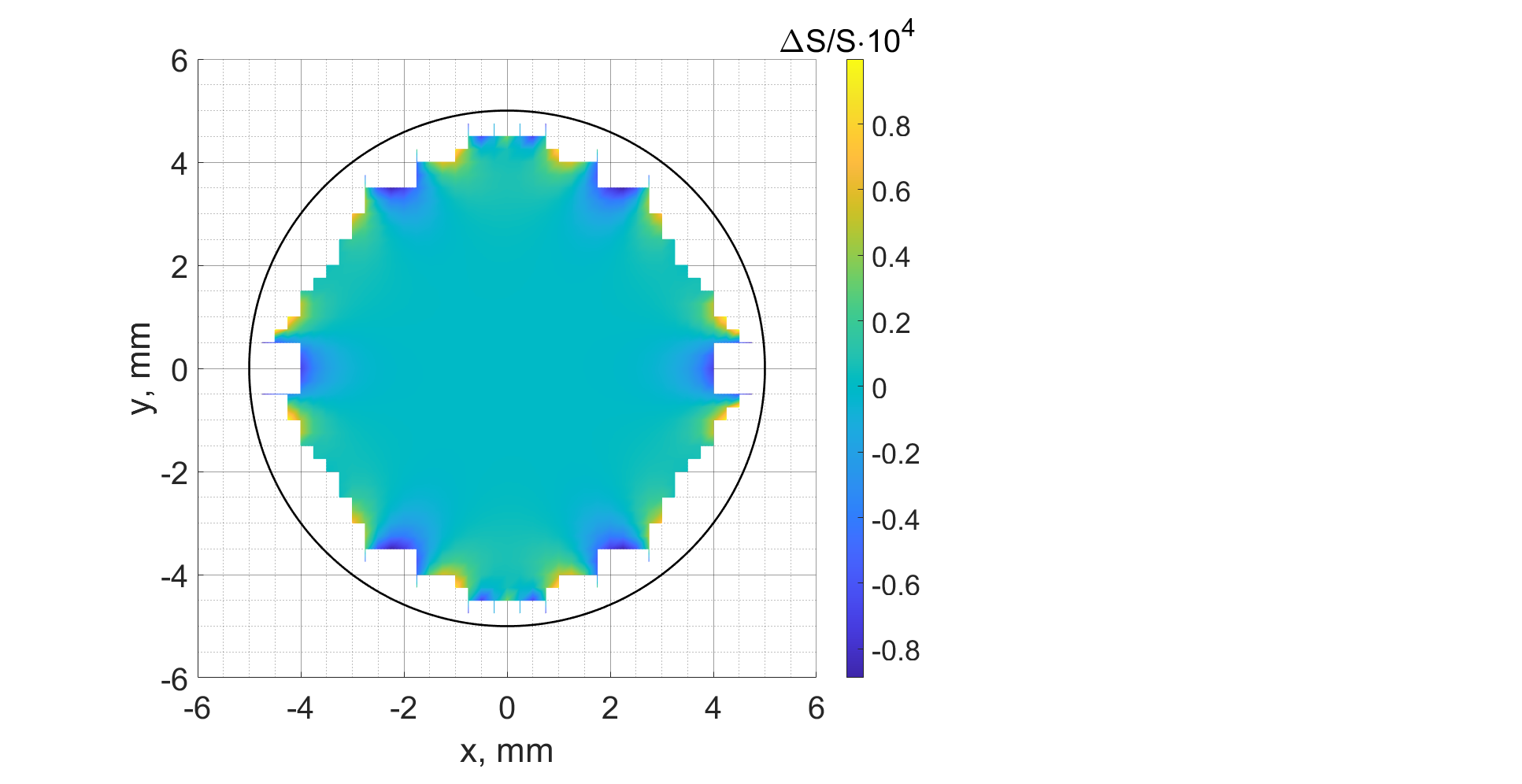}
\caption{Uniformity of the sextupole gradient (in color) of the proposed sextupole magnet with internal aperture of $R=5$~mm and $S=52\times 10^3\text{ T/m}^2$.}
\label{fig:Proposal-sextupole-3}
\end{figure}

Since we plan to vary the sextupole gradient, the sextupole should be an electromagnet. Our calculations with COMSOL Multiphysics\textsuperscript{\textregistered} software~\cite{comsol} showed that with aperture radius $R=5$~mm it is possible to manufacture the normal conducting sextupole magnet with the maximum gradient of $S=52\times 10^3\text{ T/m}^2$. In spite of saturated iron, field quality in the aperture is $10^{-4}$ with 250~A of excitation current and 12 turns per coil. The chosen length of the magnet is $L=1.59$~m. The general view, field and uniformity of the sextupolar gradient of the proposed magnet are shown on FIG.~\ref{fig:Proposal-sextupole-2} and FIG.~\ref{fig:Proposal-sextupole-3}. The magnet yoke is made of anisotropic steel 3425 to provide maximum gradient.

The neutron beam parameters are $E=10^{-4}$~eV, uniform spatial distribution with $x_0=0$~mm, $y_0=2$~mm, $\Delta x=\Delta y=0.25$~mm, 
normal angular distribution $\alpha_x=\pm0.4'=\pm1.16\times 10^{-4}$, vertical $\alpha_y=\pm1.16\times 10^{-4}$ (3 standard deviations). The corresponding period of spatial oscillations is $\lambda=1.59$~m at $S=52\times10^3\text{ T/m}^2$. Detector is placed at $L=0.1$~m from the end of the sextupole.

\begin{figure*}[htbp]
\centering
\includegraphics*[width=\columnwidth,trim=0 0 0 0, clip]{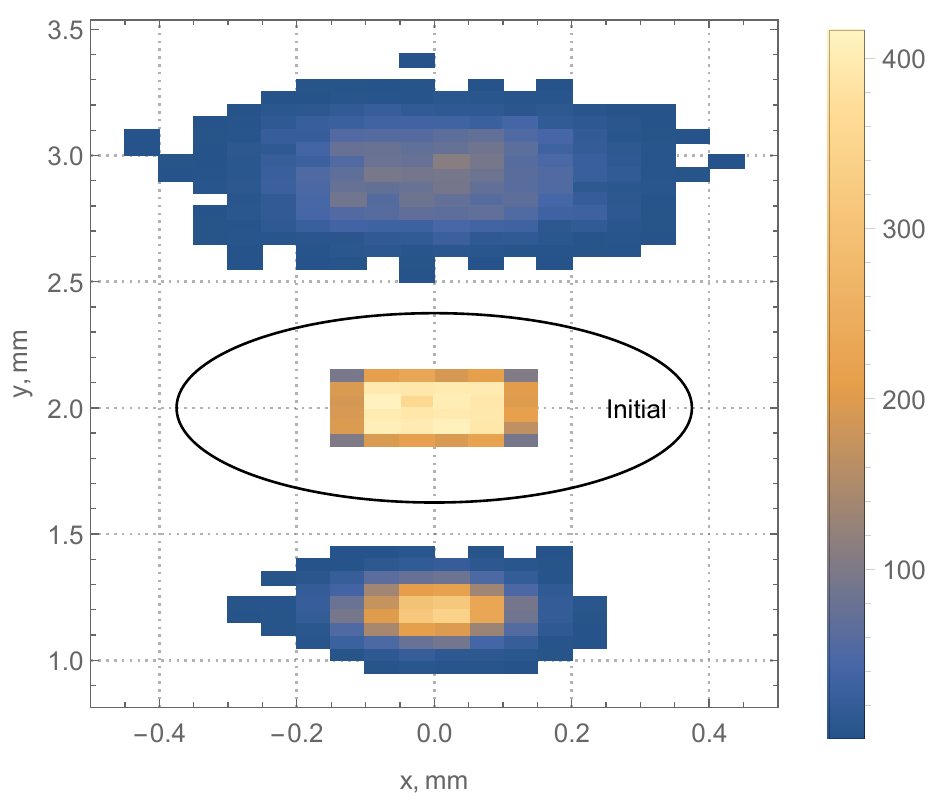}
\includegraphics*[width=\columnwidth,trim=0 0 0 0, clip]{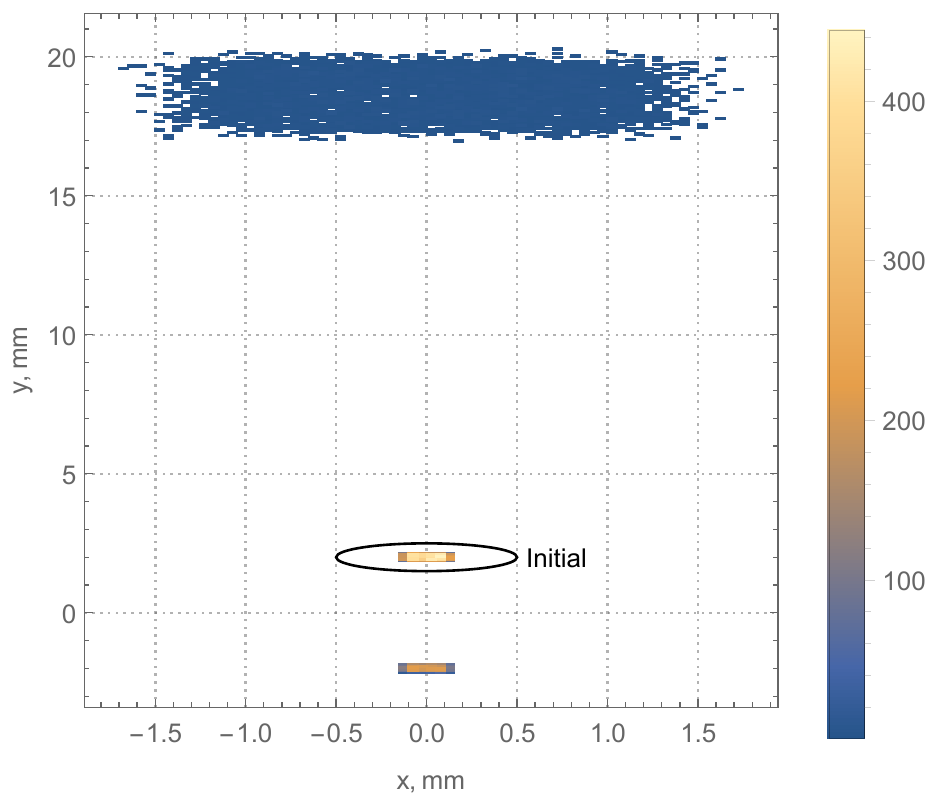}
\caption{2d coordinate beam distribution on the detector for $S=1\times 10^3\text{ T/m}^2$ (left) and  $S=1\times 10^4\text{ T/m}^2$ (right) with addition of initial beam position and distribution. The top (larger) spot corresponds to Hamiltonian \eqref{eq:ham-infin-skew}, the bottoms (smaller) corresponds to Hamiltonian \eqref{eq:ham-fin-skew}. Color denotes the number of particles.}
\label{fig:Proposal-2dxy-1e-4eV-1.59m}
\end{figure*}

The beam deflections for two sextupole strengths of $S=10^3\text{ T/m}^2$ and $S=10^4\text{ T/m}^2$, shown on FIG.~\ref{fig:Proposal-2dxy-1e-4eV-1.59m}, reveal asymmetry in deflection of the infinite and finite trajectories at higher sextupole gradient, which will indicate validity of our approach.

\section{Neutron storage ring }
Accumulation of neutrons in the storage ring will allow a number of experiments including time of life measurements with higher accuracy than before. Some experiments require focusing of the neutron beam into the smaller size, increasing the particle density. Storage ring ``Nestor'' \cite{KUGLER1985} with detailed injection system already demonstrated neutron accumulation. Neutron beam focusing by the sextupole magnets was realized in \cite{Iwashita:2024}. Now, it only remains to combine accumulation and focusing in one facility.

Construction of the storage ring with discrete magnets requires matching not only the beam trajectory with the following element, but also spin with the direction of the magnetic field.

\subsection{Beam size propagation}
For the centered beam ($\left<u\right>=0$, $\left<u'\right>=0$), the beam ``sigma'' matrix is defined by \cite{HandbookAccelerator} 
\begin{equation}
\label{eq:sigma-matrix}
\mathbf{\Sigma}_u=
\begin{pmatrix}
\left<u^2\right> & \left<u u'\right> \\
\left<u u'\right>  & \left<u'^2\right>
\end{pmatrix},
\end{equation}
where $\left<\right>$ denotes average over the beam, $u$ denotes $x$ or $y$, and expression $\mathbf{X}^T\mathbf{\Sigma}_u^{-1}\mathbf{X}=1$ (the same for $\mathbf{Y}$) describes beam ellipse in $\{u,u'\}$ plane. We consider $x$ and $y$ dimensions separately because equations of motion \eqref{eq:sext-x-motion} and \eqref{eq:sext-y-motion} are not coupled.
Using solutions \eqref{eq:sext-sol-fin} and \eqref{eq:sext-sol-inf} we find $\left<x x'\right>=0$, $\left<y y'\right>=0$, and ``sigma'' matrix \eqref{eq:sigma-matrix} for neutron becomes diagonal.

The transformation of the beam ellipse from position $1$ to position $2$ is given by
\begin{equation}
\mathbf{\Sigma}_2=\mathbf{R}\mathbf{\Sigma}_1 \mathbf{R}^{T},
\end{equation}
where $\mathbf{R}$ is a transport matrix.

\subsection{Storage ring and beam focusing}
\begin{figure}[htbp]
\includegraphics[width=\columnwidth,trim=5 0 25 0, clip]{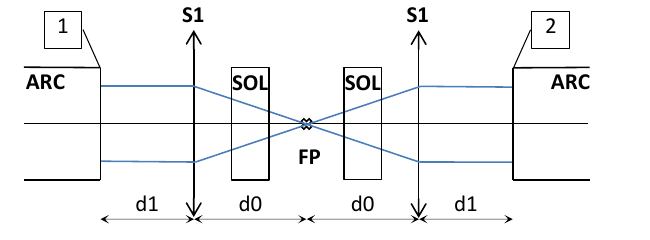}
\caption{\label{fig:telescope-2q} Sketch of interaction region: S1 -- focusing sextupoles of strength $K_1$ and length $L$,  ARC -- toroidal arc sextupole, blue lines -- beam rays, FP -- focusing point, $d_0$ and $d_1$ are the distances between the elements.}
\end{figure}

Storage ring consists of the toroidal sextupole forming a closing arc and interaction region (IR), see FIG.~\ref{fig:telescope-2q}. The telescopic transformation between the opposite ends, 1 and 2, of the closing arc is organized by two sextupoles $S1$, drifts $d_1$ and $d_0$. The transport matrix in both planes x and y is chosen to be $-I$,
\begin{equation}
\label{eq:IR-2q-telescope}
\mathbf{R}_{12}=
\begin{pmatrix}
-1 & 0 \\
0  & -1
\end{pmatrix},
\end{equation}
creating a mirror image of the beam and ensuring spin matching due to the sextupole field symmetry (expression \eqref{eq:sextupole-field-ham}  and FIG.~\ref{fig:sext-traj-xy-1}).

In order to obtain telescopic transformation \eqref{eq:IR-2q-telescope}, the lengths of the drifts should be
\begin{equation}
d_0=d_1=\frac{\cot(K_1L)}{K_1},
\end{equation}
where $K_1$ and $L$ are sextupole $S1$ strength and length respectively. The strength $K_1$ of sextupole defines the minimum beam size at the focusing point (FP) as
\begin{equation}
\label{eq:size-IP}
\left<u^2\right>_{FP}=\frac{1}{K_1^2\sin^2(K_1L)}\left<u'^2\right>_{arc},
\end{equation}
where subscript $FP$ or $arc$ denote corresponding position. Minimization of the beam size at the FP requires minimum beam angular spread at the end of the arc and large $K_1$ of the focusing sextupole, which demands smaller neutron energy and higher sextupole gradient.

The arc is a toroidal sextupole; therefore, using \eqref{eq:sext-sol-fin} and \eqref{eq:sextupole-definitions} we find relation between beam size and angular spread as 
\begin{equation}
\left<u'^2\right>_{arc}=K_{1,arc}^2\left<u^2\right>_{arc}=\frac{\left|\mu\right| \left|S_{arc}\right|M}{p_z^2}\left<u^2\right>_{arc}.
\end{equation}
Hence, the arc sextupole should have a large aperture to accumulate more particles and smaller gradient to reduce the angular spread.

\subsection{Storage ring design}
The neutron storage ring accumulates neutrons  in the toroidal sextupole making a beam occupying the whole circumference. The radius of toroid is limited by the sextupole aperture due to the transfer of the longitudinal momentum into transverse oscillations as
\begin{equation}
R_0\geq\frac{4E}{\left|\mu\right| \left|S\right|a},
\label{eq:toroid-radius}
\end{equation}
where $a$ is aperture radius of the toroidal sextupole. For neutron energy $E=10^{-6}$~eV, sextupole gradient $S=10^4$~T/m$^2$, aperture radius $a=5$~mm toroid radius is $R_0=1.3$~m.
The arc toroidal sextupole and focusing sextupoles should have the same apertures to avoid the neutron loss on geometrical aperture. 

In our conceptual design of the neutron storage ring, we have not discussed some important topics, which require detailed investigation, such as neutron injection and storage, neutron beam spatial distribution with realistic energy spread, energy acceptance of the ring, and estimations of neutron beam lifetime.

\section{Conclusion}
We solved the quantum problem of neutron motion in $2(n+1)$ pole magnet. With large quantum numbers, neutron motion in the magnetic field is reduced to classical motion of the spinless particle in two distinct potentials depending on the spin direction. In the first potential, particles experiences finite motion, and spin is parallel to magnetic field; in the second, particle's trajectory is infinite, and spin is antiparallel to magnetic field. As a result, the beam of unpolarized neutrons in the field of $2(n+1)$ pole magnet splits into two (Stern-Gerlach effect). One beam leaves the magnet according to infinite trajectory, the second beam could be trapped in the magnet with sufficient length, thus creating a neutron trap. Trajectory of the finite motion depends on initial conditions and on a $2(n+1)$ pole magnet; therefore, the trapped neutron beam will occupy the whole aperture of the magnet. However, in every trajectory point neutron beam is polarized and spin is parallel to magnetic field. This property allows to design not only the trap but also a storage ring. Conceptual design of such a neutron storage ring is presented.

\begin{acknowledgments}
We would like to express sincere gratitude to S.~Nikitin, N.~Mezentsev for fruitful discussions.
\end{acknowledgments}

\bibliography{References.bib}

\end{document}